\documentclass[11pt]{article}
\usepackage{latexsym}  
\usepackage{amssymb}
\usepackage{graphicx}
\usepackage{amsmath}

\begin{document}


\hspace{8cm}{OU-HET-709/2011; MISC-2011-09}


\begin{center}
{\Large\bf SU(5)-Compatible Yukawaon Model}

\vspace{3mm}
{\bf Yoshio Koide}

{\it Department of Physics, Osaka University,  
Toyonaka, Osaka 560-0043, Japan} \\
{\it E-mail address: koide@het.phys.sci.osaka-u.ac.jp}

\date{\today}
\end{center}

\vspace{3mm}
\begin{abstract}
It is investigated what constraints are imposed on 
the so-called yukawaon model when we regard quarks 
and leptons as $(\bar{\bf 5}_i+{\bf 10}_i+{\bf 1}_i)$
of SU(5) ($i=1,2,3$: index of U(3) family symmetry). 
In past yukawaon models, the effective Yukawa 
coupling constants $Y_f^{eff}$ are given by vacuum 
expectation values of fields $Y_f$ (``yukawaons") as 
$(Y_f^{eff})_{ij}= (y_f/\Lambda)\langle (Y_f)_{ij} \rangle$. 
In order to build a model without a cutoff scale $\Lambda$,
vector-like fields $(\bar{\bf 5}^i+{\bf 10}^i+{\bf 5}_i
+\bar{\bf 10}_i)$ are introduced in addition to 
the conventional quarks and leptons. 
The U(3) family symmetry is broken at 
$\langle Y_f\rangle \sim 10^{13}$ GeV.
\end{abstract}

\vspace{5mm}

{\large\bf 1. Introduction}

In the standard model of the quarks and leptons, the mass spectra
and mixings are due to the Yukawa coupling constants.
Then, one may investigate relations among those fundamental 
constants by assuming a family symmetry (e.g.
a U(1) symmetry, a discrete symmetry, and so on).
In contrast to such conventional models, there is another 
idea: the mass spectra and mixings are due to 
vacuum expectation values (VEVs) of new scalars.
As one of such models, the so-called ``yukawaon" model 
\cite{yukawaon} is known.

In the yukawaon model, which is a kind of 
``flavon" model \cite{flavon},  all effective 
Yukawa coupling constants $Y_f^{eff}$  ($f=u,d,e,\cdots$) 
are given by VEVs of ``yukawaons" $Y_f$ as 
$$
(Y_f^{eff})_{ij} = \frac{y_f}{\Lambda} \langle (Y_f)_{ij} \rangle ,
\eqno(1.1)
$$
where the indices $i,j$ denote 
$i,j=1,2,3$ in a family symmetry U(3).
For example, would-be Yukawa interactions are given by 
the following superpotential \cite{O3_PLB09}:
$$
W_Y = \frac{y_e}{\Lambda} {\ell}_i Y_e^{ij} e_j^c H_d 
+ \frac{y_\nu}{\Lambda} {\ell}_i Y_\nu^{ij}\nu_j^c H_u 
+\lambda_R \nu_i^c Y_R^{ij} \nu_j^c 
+ \frac{y_u}{\Lambda} u_i^{c} Y_u^{ij} q_j H_u 
+ \frac{y_d}{\Lambda} d_i^c Y_d^{ij} q_j H_d ,
\eqno(1.2)
$$
where $\ell$ and $q$ are SU(2)$_L$ doublets 
$\ell=(\nu_L, e_L)$ and $q=(u_L, d_L)$.

Note that the yukawaons $Y_f$ are singlets under the conventional
gauge symmetries SU(3)$_c \times$ SU(2)$_L\times$U(1)$_Y$,
and they have only family indices.
This suggests that the yukawaon model may be compatible with an 
SU(5) grand unification (GUT) model \cite{SU5}.
For example, we may consider: 
$$
W_Y =  \frac{y_u}{\Lambda} {\bf 10}_i Y_{(10,10)}^{ij} {\bf 10}_j {\bf 5}_H 
+ \frac{y_d}{\Lambda}\bar{{\bf 5}}_i  Y_{(5,10)}^{ij} {\bf 10}_j \bar{\bf 5}_H 
+ \frac{y_\nu}{\Lambda} \bar{{\bf 5}}_i  Y_{(5,1)}^{ij}{\bf 1}_j {\bf 5}_H 
+\lambda_R {\bf 1}_i Y_{(1,1)}^{ij} {\bf 1}_j ,
\eqno(1.3)
$$
where $\bar{{\bf 5}}+{\bf 10}+{\bf 1}$ are quark and lepton fields and
${\bf 5}_H $ and $\bar{\bf 5}_H$ correspond to the conventional 
two Higgs doublets $H_u$ and $H_d$, respectively.
However, in the would-be Yukawa interactions (1.3), the charged lepton 
yukawaon $Y_e$ is identical with the down-quark yukawaon $Y_d$, 
i.e. $Y_e=Y_d=Y_{(5,10)}$, 
although $Y_u$ and $Y_\nu$ are reasonably given by 
$Y_u =Y_{(10,10)}$ and $Y_\nu = Y_{(5,1)}$. 
In the yukawaon model, it is essential that $\langle Y_e \rangle$ has 
a different family structure from $\langle Y_d \rangle$.
We must build a model where yukawaon $Y_e$ is an independent field
from $Y_d$.
For this problem, we will introduce matter fields $f^{\prime}
=(\bar{\bf 5}'_i+{\bf 5}^{\prime\, i})$ in addition to the quarks and leptons
$f_i =(\bar{\bf 5}+{\bf 10}+{\bf 1})_i$.

The purpose of the present paper is to investigate what constraints 
are imposed on the so-called yukawaon model when we regard quarks 
and leptons as $(\bar{\bf 5}_i+{\bf 10}_i+{\bf 1}_i)$ of SU(5). 
In this paper, we do not try to unify a family symmetry 
into the SU(5) gauge symmetry.
We do not intend to resolve current problems in the minimum SU(5) 
GUT by considering the yukawaon model.
Since yukawaons are SU(5) singlets, the existence of the yukawaons
do not affect an SU(5) GUT model, so that we can inherit the successful 
results in the SU(5) GUT and we also inherit the current problems in the 
minimum SU(5) GUT scenario. 
For example, it is well known that it is difficult to build an SU(5) 
GUT model in which $R$ charges are conserved.
However, in the yukawaon model, $R$ charges have to be conserved in 
order to distinguish yukawaons $Y_f$ from the each other. 
In the present model, too, we assume $R$ charge conservation,  
while we do not touch whether this is compatible with SU(5) GUT scenario
or not.
(Of course, $R$ charge conservation is broken below $\mu = \Lambda_{fam}$,
where the family symmetry U(3) is broken at $\mu \sim \Lambda_{fam}$.)  
We optimistically consider that the problem will be resolved 
in a scenario with a higher GUT group and/or with 
an extra-dimension.
In the present paper, we pay attention only to yukawaon sector.

So far, as seen in Eq.(1.2), the yukawaon model has been based on
an effective theory with a cutoff $\Lambda$.
Another purpose of the present SU(5) compatible model is to make the 
meaning of the cutoff scale $\Lambda$ clear by considering a model
without such a cutoff scale. 
In order to build a model without a cutoff scale $\Lambda$, 
we will introduce matter fields 
$f^{\prime\prime}=( \bar{\bf 5}^{\prime\prime}+{\bf 10}^{\prime\prime})^i 
+ ({\bf 5}^{\prime\prime}+\overline{\bf 10}^{\prime\prime})_i$
in addition to the fields $f_i$ and $f^{\prime}$.
(Note that $f^{\prime\prime}$ do not mix with the quarks and leptons
$f_i$ because quarks and leptons $f_i=(\bar{{\bf 5}}(\bar{{\bf 5}}')+{\bf 10}
+{\bf 1})_i$ are ${\bf 3}$ of U(3), while $( \bar{\bf 5}^{\prime\prime}
+{\bf 10}^{\prime\prime})^i $ are ${\bf 3}^*$ of U(3). )

As we discuss later, the existence of the new matter fields
$f'$ and $f^{\prime\prime}$ will bring some merits and demerits in 
the yukawaon model: 
(i) $Y_\nu$ in Eq.(1.2) will bring  naturally substituted with
$Y_e$ in the present model, although the replacement $Y_\nu \rightarrow Y_e$ 
was an ad hoc ansatz in the previous yukawaon model \cite{O3_PLB09}; 
(ii) In the past yukawaon model, in order to explain the observed 
fact $m_t \sim \langle H_u \rangle$, we must consider 
$\langle Y_u \rangle/\Lambda \sim 1$ (we consider $\tan\beta \sim 10$).
This is somewhat controversial within the framework of such 
an effective theory with $\Lambda$. 
In contrast to the old model, in the present model, the factor 
$\langle Y_u \rangle/\Lambda$ in the previous model corresponds to 
$\langle Y_u \rangle/M_{10}$ ($M_{10}$ is a mass parameter in the 
superpotential and it is given by 
$M_{10} \overline{\bf 10}_i^{\prime\prime} {\bf 10}^{\prime\prime\, i}$
), and we can reasonably take $M_{10} \sim \langle Y_u \rangle$.
(iii) It is difficult to lower a scale $\Lambda_{fam}$  of the U(3) 
family symmetry breaking in the present SU(5) compatible model, 
in order that the gauge coupling constants of SU(3)$_c \times$SU(2)$_L 
\times$U(1)$_Y$ do not blow up under the additional particles 
$f^{\prime\prime}=( \bar{\bf 5}^{\prime\prime}+{\bf 10}^{\prime\prime})^i 
+ ({\bf 5}^{\prime\prime}+\bar{\bf 10}^{\prime\prime})_i$.

In the next section, we investigate possible would-be Yukawa 
interactions without a cutoff parameter $\Lambda$ and 
an answer to the above question (ii) will be given. 
In Sec.3, we investigate superpotential for the yukawaon sector.
Although the VEV relations are substantially inherited from 
previous those, the superpotential forms are considerably 
changed because we do not consider terms which include a cutoff
scale $\Lambda$. 
In Sec.4, we discuss a possible scale of $\Lambda_{fam}$ at 
which U(3) family symmetry is broken.
We will conclude $\Lambda_{fam} \sim 10^{13}$ GeV.
Sec.5 is devoted to concluding remarks.

\vspace{5mm}

{\large\bf 2. Would-be Yukawa interactions}

First, le us discuss the $Y_e$-$Y_d$ splitting.
We introduce vector-like 
${\bf 5}^{\prime\, i}$ and $\bar{\bf 5}'_i$ fields in addition to 
the fields given in Eq.(1.3). 
For convenience, we denote one ${\bf 5}$ and 
two $\bar{\bf 5}$ as
$$
\bar{\bf 5}_i = (D_i^c, \ell_i), \ \ \ 
\bar{\bf 5}'_i = (d_i^c, L_i), \ \ \ 
{\bf 5}^{\prime\, i} =( \bar{D}^{c\, i}, \bar{L}^{i}),
\eqno(2.1)
$$
where $d^c$, $D^c$ and $\bar{D}^c$ are SU(2)$_L$ singlet 
down-quarks with electric charges $+1/3$, $+1/3$
and $-1/3$, respectively, and $\ell$, $L$ and $\bar{L}$ are
SU(2)$_L$ lepton doublets.
In order to realize that the fields $(\bar{D}^c, D^c)$, 
and $(\bar{L}^c, L^c)$ become massive and decouple from 
the present model, we assume the following interactions
$$
\lambda_D \bar{\bf 5}_i^A(\Sigma_3)_A^B {\bf 5}_B^{\prime\, i} 
+ \lambda_L (\bar{\bf 5}')_i^A(\Sigma_2)_A^B {\bf 5}_B^{\prime\, i} ,
\eqno(2.2)
$$
where indices $A,\ B=1,2, \cdots, 5$ are those of SU(5),
and SU(5) ${\bf 24}+{\bf 1}$ fields $\Sigma_2$ and $\Sigma_3$ 
take VEV forms 
$$
\begin{array}{l}
\langle \Sigma_2 \rangle = v_2 \, {\rm diag}(0,0,0,1,1)  , \\ 
\langle \Sigma_3 \rangle =  v_3\, {\rm diag}(1,1,1,0,0) . 
\end{array}
\eqno(2.3)
$$
Therefore, Eq.(2.2) leads to mass terms
$$
\lambda_D v_{3} \bar{D}^{c\, i} D^c_i + \lambda_L v_{2} \bar{L}^i L_i .
\eqno(2.4)
$$
As we discuss later, we will assign different $R$ charges to 
$\Sigma_2$ and $\Sigma_3$ (and also to ${\bf 5}$ and ${\bf 5}'$). 
If we once admit the existence of $\Sigma_3$ with
VEV given in Eq.(2.3) phenomenologically, we can use 
$\Sigma_3$ for the triplet-doublet splitting of the Higgs
fields $\bar{\bf 5}_H$ and ${\bf 5}_H$ by considering 
an interaction $\bar{\bf 5}_H \Sigma_3 {\bf 5}_H$.
For the triplet-doublet splitting in Higgs fields, 
already a reasonable mechanism has been proposed based on
SO(10) GUT scenario \cite{T-D_splitting}.
Since the purpose of the present paper is not to investigate 
the GUT scenario, we ad hoc assume the VEV forms (2.3) 
in this paper.
The reason of the forms (2.3) may be understood by a scenario 
based on a higher gauge GUT group and/or on extra-dimensions. 

Next, we discuss a seesaw-type mass matrix mechanism 
in order to realize the effective interactions given by 
Eq.(1.2).
We assume further vector-like matter fields $f^{\prime\prime}
=(\bar{\bf 5}^{\prime\prime} + {\bf 10}^{\prime\prime})^i 
+ ({\bf 5}^{\prime\prime} + \overline{\bf 10}^{\prime\prime})_i$. 
The up-quark, charged lepton, and down-quark mass matrices, 
$M_u$, $M_e$ and $M_d$,  
are generated by the following superpotential:
$$
W_u = y_{u} {\bf 10}_i Y_u^{ij} \overline{\bf 10}_j^{\prime\prime} 
+M_{10} \overline{\bf 10}_i^{\prime\prime}  {\bf 10}^{\prime\prime\, i}
+ y_{10} {\bf 10}^{\prime\prime\, i} {\bf 10}_i {\bf 5}_H ,
\eqno(2.5)
$$
$$
W_{e,d} = y_{e} \bar{\bf 5}_i Y_e^{ij} {\bf 5}_j^{\prime\prime} 
+ y_{d} \bar{\bf 5}'_i Y_d^{ij} {\bf 5}_j^{\prime\prime} 
+M_{5} {\bf 5}_i^{\prime\prime} \bar{\bf 5}^{\prime\prime\, i} 
+ y_{5} \bar{\bf 5}^{\prime\prime\, i} {\bf 10}_i \bar{\bf 5}_H ,
\eqno(2.6)
$$
which lead to effective Yukawa interactions
$$
W_u^{eff} =\frac{y_u y_{10}}{\bar{M}_{10}^{(ij)}} {\bf 10}_i Y_{u}^{ij} 
{\bf 10}_j {\bf 5}_H ,
\eqno(2.7)
$$
$$
W_{e,d}^{eff} = \frac{y_e y_{5}}{\bar{M}_{5e}^{(ij)}} 
\bar{\bf 5}_i Y_{e}^{ij} {\bf 10}_j \bar{\bf 5}_H 
+ \frac{y_d y_{5}}{\bar{M}_{5d}^{(ij)}}  \bar{\bf 5}'_i Y_{d}^{ij} 
{\bf 10}_j \bar{\bf 5}_H ,
\eqno(2.8)
$$
respectively.
Note that the cutoff parameters have flavor-dependence
differently from the conventional seesaw model, 
because we are interested in cases of 
$\langle Y_u^{33} \rangle/M_{10} \sim 1$ and 
$\langle Y_{e,d}^{33} \rangle/M_{5} \sim 10^{-1}$.
Here, the factors $\bar{M}_{10}^{(ij)}$, $\bar{M}_{5e}^{(ij)}$ 
and $\bar{M}_{5d}^{(ij)}$ are given by
$$
\bar{M}_{10}^{(ii)} \simeq M_{10} \left[1 +
(\langle Y_u^{ii} \rangle/M_{10})^2 \right] , 
\eqno(2.9)
$$
$$
\bar{M}_{5e}^{(ii)} \simeq \sqrt{(M_{5})^2 +
(\langle Y_e^{ii} \rangle)^2 }, \ \  
\bar{M}_{5d}^{(ii)} \simeq \sqrt{ (M_{5})^2 +
(\langle Y_d^{ii} \rangle)^2 } ,
\eqno(2.10)
$$
in the bases in which the VEV matrices $\langle Y_u \rangle$,
$\langle Y_e \rangle$ and $\langle Y_d \rangle$ are 
diagonal, respectively.
The expressions (2.9) and (2.10) are obtained from 
diagonalizing the mass matrices for $({\bf 10}, {\bf 10}^{\prime\prime},
\overline{\bf 10}^{\prime\prime})$ and 
$(\bar{\bf 5} \ (\bar{\bf 5}'), {\bf 10}, \bar{\bf 5}^{\prime\prime},
{\bf 5}^{\prime\prime})$:
$$
M_{10}^{u} = \frac{1}{2} \left(
\begin{array}{ccc}
0 &  y_{10} v_{Hu} & y_u \langle Y_u \rangle \\
y_{10} v_{Hu} & 0 & M_{10} \\
y_u \langle Y_u \rangle & M_{10} & 0
\end{array} \right) , \ \ \ 
M_{5}^{e,d} = \left(
\begin{array}{cccc}
0 & 0 & 0  & y_{e,d} \langle Y_{e,d} \rangle \\
0 & 0 & y_{5} v_{Hd} & 0 \\
0 & y_{5} v_{Hd} & 0 & M_5 \\
y_{e,d} \langle Y_{e,d} \rangle & 0 & M_5 & 0
\end{array} \right) , \ \ \ 
\eqno(2.11)
$$
respectively, where 
$v_{Hu} = \langle H_u^0 \rangle=\langle {\bf 5}_H\rangle$ 
and 
$v_{Hd} = \langle H_d^0 \rangle=\langle \bar{\bf 5}_H\rangle$.

For a neutrino Dirac mass term, we also assume 
$$
W_\nu = y_{e} \bar{\bf 5}_i Y_e^{ij} {\bf 5}_j^{\prime\prime} 
+M_{5} {\bf 5}_i^{\prime\prime} \bar{\bf 5}^{\prime\prime\, i} 
+ y_{1} \bar{\bf 5}^{\prime\prime\, i} {\bf 1}_i {\bf 5}_H ,
\eqno(2.12)
$$
which leads to the effective interaction
$$
W_\nu^{eff} = \frac{y_e y_{1}}{M_{5}} \bar{\bf 5}_i Y_{e}^{ij} 
{\bf 1}_j {\bf 5}_H .
\eqno(2.13)
$$
(In the present model, we do not assume ${\bf 1}_i^{\prime\prime}+
{\bf 1}^{\prime\prime\, i}$.)
Note that the neutrino Dirac mass matrix $m_D$ is proportional
to $\langle Y_e \rangle$ as well as the charged lepton mass
matrix $M_e$.
When we assume that the SU(5) singlet matter field ${\bf 1}_i$
can have a Majorana mass term
$$
W_R = \lambda_R {\bf 1}_i Y_R^{ij} {\bf 1}_j ,
\eqno(2.14)
$$
we can obtain a seesaw-type neutrino mass matrix
$$
M_\nu = \frac{y_e^2 y_1^2}{\lambda_R} \left( 
\frac{v_{Hu}}{M_5} \right)^2 
\langle Y_e \rangle \langle Y_R \rangle^{-1} 
\langle Y_e \rangle ,
\eqno(2.15)
$$
under an approximation $\bar{M}_{5e} \simeq M_5$, 
so that we can rewritten Eq.(2.15) as
$$
M_\nu = \frac{y_1^2 \tan^2 \beta}{\lambda_R y_5^2}  \, 
M_e \, \langle Y_R \rangle^{-1} M_e ,
\eqno(2.16)
$$
where $ \tan\beta =\langle H_u^0 \rangle/\langle H_d^0 \rangle$.
By taking $m_\tau \simeq 1.777$ GeV, $(M_{\nu })_{33}
\sim m_{\nu 3} \simeq \sqrt{\Delta m_{atm}^2} \simeq 0.049$ eV 
\cite{PDG10}) and $\tan\beta \simeq 10$,
we can estimate the value of $\langle Y_R \rangle$ as
$$
\langle Y_R \rangle  \simeq \lambda_R (y_5/y_1)^2 \times 6.4
\times 10^{12} \ {\rm GeV}.
\eqno(2.17)
$$

$R$ charges of these matter fields must satisfy the following 
relations:
$$
R({\bf 10}_i) - R({\bf 1}_i) = 
R(\bar{\bf 5}_i^{\prime\prime}) - R({\bf 10}_i^{\prime\prime}) = 
R(\overline{\bf 10}^{\prime\prime\, i}) - R({\bf 5}^{\prime\prime\, i}) = 
R({\bf 5}_H) - R(\bar{\bf 5}_H) \equiv a ,
\eqno(2.18)
$$
$$
R(\bar{\bf 5}'_i) - R(\bar{\bf 5}_i) =
R(\Sigma_3) - R(\Sigma_2) \equiv b ,
\eqno(2.19)
$$
$$
R(\bar{\bf 5}^{\prime\prime\, i}) + R({\bf 5}^{\prime\prime}_i) =2 ,
\eqno(2.20)
$$
$$
R(\bar{\bf 5}_i) + R({\bf 5}^{\prime\, i}) +
R(\Sigma_3) = 2 ,
\eqno(2.21)
$$
$$
R(Y_R) + 2 R({\bf 1}_i) =2 .
\eqno(2.22)
$$
Since there are still free parameters in these assignments, 
we do not show an explicit assignment of $R$ charges in
this paper.

\vspace{5mm}

{\large\bf 3. Superpotential for yukawaons}

The successful results in the previous 
yukawaon model with an O(3) family symmetry \cite{O3_PLB09} have 
been derived on the basis of the following VEV relations:
$$
\langle Y_e \rangle = k_e \langle \Phi_e \rangle 
\langle \Phi_e \rangle,
\eqno(3.1)
$$
$$
\langle Y_u \rangle = k_u \langle \Phi_u \rangle 
\langle \Phi_u \rangle,
\eqno(3.2)
$$ 
$$
\langle \Phi_u \rangle = k'_u \langle \Phi_e \rangle 
(\langle E \rangle + a_u \langle X \rangle ) \langle \Phi_e \rangle,
\eqno(3.3)
$$
$$
\langle Y_d \rangle = k_d \langle \Phi_e \rangle 
(\langle E \rangle + a_u \langle X \rangle ) \langle \Phi_e \rangle,
\eqno(3.4)
$$
$$
\langle Y_R \rangle = k_R \left[ \langle \Phi_u \rangle 
\langle P_u \rangle \langle Y_e \rangle +
\langle Y_e \rangle \langle P_u \rangle 
\langle \Phi_u \rangle  + \xi_\nu (
\langle \Phi_u \rangle \langle Y_e \rangle 
\langle P_u \rangle + \langle P_u \rangle
\langle Y_e \rangle \langle \Phi_u \rangle )
\right] , 
\eqno(3.5)
$$
where 
$$
\langle \Phi_e \rangle_e \propto {\rm diag}
(\sqrt{m_e}, \sqrt{m_\mu}, \sqrt{m_\tau}) ,
\eqno(3.6)
$$
$$
\langle E \rangle_e = v_E \left( 
\begin{array}{ccc}
1 & 0 & 0 \\
0 & 1 & 0 \\
0 & 0 & 1
\end{array} \right) , \ \ 
\langle X \rangle_e = \frac{1}{3} v_X \left( 
\begin{array}{ccc}
1 & 1 & 1 \\
1 & 1 & 1 \\
1 & 1 & 1
\end{array} \right) , \ \ 
\langle P_u \rangle_u = v_{Pu} \left( 
\begin{array}{ccc}
1 & 0 & 0 \\
0 & -1 & 0 \\
0 & 0 & 1
\end{array} \right) .
\eqno(3.7)
$$ 
Those VEV relations have been derived from 
superpotential by using SUSY vacuum 
conditions, so that the relations are dependent 
on the quantum number assignments of the yukawaons.
In other words, the relations are based on completely 
phenomenological assumptions (quantum number assignments).
The bilinear form of $Y_e$ in Eq.(3.1) is required to
give a charged lepton mass relation (see Eq.(3.9) later).
The bilinear form of $Y_u$ in Eq.(3.2) plays an 
inevitable role in obtaining a successful neutrino mass 
matrix via Eq.(3.5).

Note that the VEV matrix forms (3.6) and (3.7) are 
dependent on the flavor basis.
The expression $\langle A \rangle_f$ ($f=e,u$) denotes 
a form of the VEV matrix $\langle A \rangle$ in a basis
in which the mass matrix $M_f$ is diagonal.  
In this paper, we do not give a superpotential which 
leads to the reasonable eigenvalues (3.6), and we 
will use the observed running mass values at $\mu = m_Z$ 
of the charged lepton masses. 
The VEV forms given in Eq.(3.7) are essentially based on
phenomenological assumptions as we state later.

Since $Y_e$ was substituted for $Y_\nu$ in the previous model
 \cite{O3_PLB09},  
the neutrino mass matrix $M_\nu$ was given by the form (2.15).
Therefore, the observed neutrino mixing originates only in 
the structure of $\langle Y_R \rangle$.
When a reasonable structure of $\langle Y_R \rangle$ is assumed, 
the model can give excellent agreement with the 
nearly tribimaximal mixing \cite{tribi} together with the
up-quark mass ratios only by adjusting two parameters [$a_u$ 
in Eq.(3.3) and $\xi_\nu$ in Eq.(3.5)]. 
Besides, the model can give rough agreement with the 
Cabibbo-Kobayashi-Maskawa (CKM) mixing parameters and 
down-quark mass ratios
by adjusting remaining two parameters [the magnitude and 
phase of $a_d$ in Eq.(3.4)].

In the previous model \cite{O3_PLB09}, these VEV relations 
(3.1) -- (3.5) have been derived from the SUSY vacuum conditions 
by assuming an O(3) family symmetry. 
However, the superpotential terms have (1.2) included a cutoff 
parameter $\Lambda$.
As seen in the previous section, in the present model, 
we want to build a model without such a cutoff parameter $\Lambda$.
Moreover, in the present paper, we will investigate a model with a 
U(3) family symmetry instead of the O(3) family symmetry, 
because, in the yukawaon model, the order of the VEV matrices
is important [e.g. absence of $(E+aX)\Phi_e \Phi_e +\Phi_e \Phi_e
(E+aX)$ in contrast to the existence of $\Phi_e (E+a X) \Phi_e$].

Recently, Sumino has proposed a charged lepton mass matrix model
based on a U(3) gauge family symmetry 
\cite{Sumino09}.
In the Sumino model, the charged lepton mass term is 
generated by a would-be Yukawa interaction 
$$
H_e = \frac{y_e}{\Lambda^2} 
\bar{\ell}_{L}^i \Phi^e_{i \alpha}
 \Phi^{eT}_{\alpha j} e_{R}^j H ,
\eqno(3.8)
$$
where $i$ and $\alpha$ are indices of U(3) and O(3), respectively,
and $H$ is the Higgs scalar in the standard non-SUSY model.
(Sumino' model has not been based on a SUSY scenario.)
The charged lepton masses $m_{e_i}$ are acquired from  
a VEV of the scalar $\Phi^e$ \cite{Koide90MPL},
i.e. the masses $m_{ei}$ are given by $m_{ei}=(y_e/\Lambda^2) 
\langle \Phi^e_{i\alpha} \rangle \langle \Phi^{eT}_{\alpha i} 
\rangle \langle H^0 \rangle$.
Sumino's interest was in the charged 
lepton mass relation~\cite{Koidemass}
$$
K \equiv \frac{m_e +m_\mu + m_\tau}{
   (\sqrt{m_e} + \sqrt{m_\mu} + \sqrt{m_\tau})^2} 
   = \frac{2}{3} . 
\eqno(3.9)
$$
The relation $K=2/3$ is satisfied with the order of $10^{-5}$ 
for the pole masses, i.e. 
$K^{pole}=(2/3)\times (0.999989 \pm 0.000014)$ 
\cite{PDG10}, while it is only valid with 
the order of $10^{-3}$ for the running masses, 
e.g. $K(\mu)=(2/3)\times (1.00189 \pm 0.00002)$ 
at $\mu =m_Z$. 
However, in conventional mass matrix models, ``mass" 
means not  ``pole mass" but ``running mass".
Sumino has seriously taken why the mass formula 
$K=2/3$ is so remarkably satisfied with the pole masses.  
The deviation of $K(\mu)$ from $K^{pole}$ is 
caused by a logarithmic term $m_{ei}\log(\mu/m_{ei})$ 
in the radiative correction term \cite{Arason} due to photon
$$
m_{ei}(\mu) = m_{ei}^{pole} \left[ 1-\frac{\alpha(\mu)}{\pi}
\left(1 +\frac{3}{4} \log \frac{\mu}{m_{ei}(\mu)} \right)
\right].
\eqno(3.10)
$$
Therefore, he has assumed that a family symmetry is local, 
and that the logarithmic term in the 
radiative correction due to photon is  canceled by 
that due to  family gauge bosons. 
In the Sumino model, it is essential that the left- and right-handed
charged leptons $e_{Li}$ and $e_{Ri}$ are assigned to ${\bf 3}$
and ${\bf 3}^*$ of U(3) family symmetry, respectively. 
(A similar fermion assignment has been proposed by 
Applequist, Bai and Piai \cite{Appelquist06}.)
Also, in his model, it is essential that the family gauge boson 
masses $m(A_i^j)$ are given by $m^2(A_i^j) \propto 
\langle \Phi_e^{i\alpha} \rangle \langle \Phi_e^{T \alpha j}\rangle 
\propto m_{ei}+m_{ej} $.  
As a result, we can obtain $K(\mu) =K^{pole}$. 
[However, it does not mean $m_{ei}(\mu)=m_{ei}^{pole}$.
The cancellation takes place only for the term with 
$\log m_{ei}$ in Eq.(3.10).] 

Stimulated by the Sumino model, in this paper, we
assume U(3)$\times$O(3) family symmetries \cite{U3O3_JPG11}.
The VEV relations (3.1) and (3.2) are derived from 
the following superpotential terms: 
$$
W_e = \mu_e (Y_e^{ij} \Theta_{ji}^e) +\lambda_e 
(\Phi_e^{i\alpha} \Phi_e^{T \alpha j} \Theta_{ji}^e),
\eqno(3.11)
$$
$$
W_u = \lambda_u [Y_u^{ik} E^u_{kj} (\Theta_u)^j_i] 
+\lambda'_u 
[(\Phi_u)^i_k (\Phi_u)^k_j (\Theta_u)^j_i] ,
\eqno(3.12)
$$
where $E^u$ takes a VEV form $\langle E^u \rangle
=v_E \, {\rm diag}(1,1,1)$.
Here and hereafter, we denote fields whose VEV values
are zeros as $\Theta_A$ ($A=e, u, \cdots$).
(Therefore, we can obtain meaningful VEV relations from
SUSY vacuum conditions $\partial W/\partial \Theta_A=0$, 
while we cannot obtain any relations 
from other conditions (e.g. $\partial W/\partial Y_f=0$) 
because the relations always include $\langle \Theta_A \rangle$.
For the time being, we assume that the supersymmetry breaking 
is induced by a gauge mediation mechanism (not including family 
gauge symmetries), 
so that our VEV relations among yukawaons are still valid 
even after
the SUSY was broken in the quark and lepton sectors.
In Eqs.(3.11) and (3.12), according to Sumino, 
we assume \cite{Sumino09} that the gauge symmetry O(3) is already  
broken at $\mu=\Lambda_{GUT}$.

Note that, for the VEV relations (3.3) -- (3.5), we cannot 
use the similar prescriptions
which were used for Eqs.(3.11) and (3.12), 
because we want 
a model without a cutoff $\Lambda$.
We assume the following superpotential forms 
without $\Lambda$ for the 
VEV relations (3.3) and (3.4):
$$
W'_u =  \lambda_u^{\prime\prime} (\Phi_u)^{i}_k P_u^{kj} \Theta^{u\prime}_{ji}
+\lambda_u^{\prime\prime\prime}\left[ \Omega_E^{i\alpha} 
\Omega_E^{T \alpha j} + a_u \Omega_X^{i\alpha} \Omega_X^{T \alpha j}\right]  \Theta^{u\prime}_{ji} ,
\eqno(3.13)
$$
$$
W_d =\mu_d Y_d^{ij} \Theta^d_{ji} + \lambda_d 
\left[ \Omega_E^{i\alpha} \Omega_E^{T \alpha j} + a_d 
\Omega_X^{i\alpha} \Omega_X^{T \alpha j} \right]  \Theta^d_{ji} ,
\eqno(3.14)
$$
together with
$$
W_{E,X} = \mu_E \Omega_E^{i \alpha} \Theta^E_{\alpha i} 
+ \lambda_E \Phi_e^{i \beta} E_\beta^\alpha \Theta^E_{\alpha i} 
+ \mu_X \Omega_X^{i \alpha} \Theta^X_{\alpha i} 
+ \lambda_X \Phi_e^{i \beta} X_\beta^\alpha \Theta^X_{\alpha i} .
\eqno(3.15)
$$
where $E$, $X$ and $P_u$ have the VEV forms defined in Eq.(3.7). 
For such VEV forms of $X$ and $E$, for example, we may consider
the following superpotential forms \cite{U3O3_JPG11}:
$$
W_X =\lambda_X \, {\rm det}X \equiv \lambda_X \left(  
\frac{1}{3} {\rm Tr}[XXX] -\frac{1}{2} {\rm Tr}[XX] {\rm Tr}[X]
+\frac{1}{6} ({\rm Tr}[X])^3 \right) ,
\eqno(3.16)
$$  
$$
W_E = \lambda_E {\rm Tr}[EE] {\rm Tr}[E] 
+\lambda'_E({\rm Tr}[E] )^3 .
\eqno(3.17)
$$
(Eq.(3.17) can uniquely lead to $\langle E \rangle \propto
{\bf 1}$, while Eq.(3.16) can only show that $\langle X\rangle$ 
is a rank 1 matrix.  It is still an ad hoc assumption  
as well as in the previous models \cite{O3_PLB09,U3O3_JPG11}
that the rank 1 matrix takes the democratic form given by 
Eq.(3.7) in the diagonal basis of $\langle \Phi_e \rangle$. )

For the VEV relation (3.5), we assume the superpotential
$$
W_R = \mu_R Y_R^{ij} \Theta^R_{ji} +
\lambda'_R [ (\Phi_u)^i_k Y_e^{kj} + Y_e^{ik} (\Phi_u)^j_k 
+\xi_\nu (\Phi_u)^k_k Y_e^{ij} ] \Theta^R_{ji} .
\eqno(3.18)
$$
Here, the final term is somewhat different from the previous form (3.5).
In the O(3) model, a term $(\Phi_u Y_e P_u + P_u Y_e \Phi_u)$ 
is allowed in addition to $(\Phi_u P_u Y_e + Y_e P_u \Phi_u)$,
because these fields are $({\bf 5}+{\bf 1})$ of O(3).
However, in the present model, such a term is forbidden.
Instead, we have added 
${\rm Tr}[\Phi_u] \, Y_e^{ij}$ to the term 
$(\Phi_u Y_e +Y_e \Phi_u)^{ij}$ as a new $\xi_\nu$ term
in Eq.(3.18). 
This term is required \cite{U3O3_JPG11} in order to fit 
the observed value of $\tan^2 \theta_{solar}$.

Thus, we can obtain the neutrino mass matrix Eq.(2.15) with
$$
\langle Y_R \rangle = -\frac{\lambda_R}{\mu_R} \left(
\langle \Phi_u \rangle \langle Y_e \rangle + 
\langle Y_e \rangle \langle \Phi_u \rangle + \xi_\nu
{\rm Tr}[\langle \Phi_u \rangle] \langle Y_e \rangle 
\right) ,
\eqno(3.19)
$$
$$ 
\langle \Phi_u \rangle \langle P_u \rangle = 
-\frac{\lambda_u^{\prime\prime\prime}}{
\lambda_u^{\prime\prime}} \left( 
\langle \Omega_E \rangle \langle \Omega_E^T \rangle +
a_u \langle \Omega_X \rangle \langle \Omega_X^T \rangle
\right)  = -\frac{\lambda_u^{\prime\prime\prime}\lambda_X^2}{
\lambda_u^{\prime\prime}\mu_X^2} v_X 
\langle \Phi_e \rangle (\langle E \rangle +a_u 
\langle X \rangle )\langle \Phi_e \rangle .
\eqno(3.20)
$$
where, for simplicity, we have put $\lambda_E/\mu_E=
\lambda_X/\mu_X$ and $v_E=v_X$. 
Numerical results for neutrino mixing parameters and
up-quark mass rations are identical with those given
in Ref.\cite{U3O3_JPG11}.
We quote the numerical results from Ref.\cite{U3O3_JPG11}
in Table 1.
Here, we have taken a value $a_u=-1.78$ which can give
reasonable up-quark mass ratios:
$$
\sqrt{\frac{ {m_u}}{{m_c}}} = 0.04389, \ \ \ \   
\sqrt{\frac{ {m_c}}{m_t}} = 0.05564.
\eqno(3.21)
$$
The predicted values (3.21) are in good agreement with 
the observed values at $\mu=m_Z$ \cite{q-mass} 
$\sqrt{ {m_u}/{m_c}} = 0.045^{+0.013}_{-0.010}$ and  
$\sqrt{ {m_c}/{m_t}} = 0.060 \pm 0.005$. 

\begin{table}
\begin{center}
\begin{tabular}{cccc} \hline
$\xi_\nu$ & $\tan^2 \theta_{solar}$ & $\sin^2 2\theta_{atm}$
& $|U_{13}|^2$ \\ \hline
$0$    & $0.6995$  & $0.9872$ & $1.72\times 10^{-4}$ \\
$0.009$ & $0.4610$ & $0.9902$ & $2.28\times 10^{-4}$ \\
$0.010$ & $0.4408$ & $0.9905$ & $2.35\times 10^{-4}$ \\
\hline
\end{tabular}  
\end{center}
\begin{quotation}
\caption{
$\xi_\nu$ dependence of the neutrino parameters.
The value of $a_u$ is taken as $a_u=-1.78$
which can give reasonable up-quark mass ratios.
}
\end{quotation}
\end{table}

In Eqs.(3.13) and (3.14), we have considered somewhat 
unbalanced assignments between $\Phi_e$ and $\Phi_u$, 
i.e. $\Phi_e^{i \alpha} =(3,3)$ and 
$(\Phi_u)^i_j =(8+1,1)$ of U(3)$\times$O(3).
We may consider an alternative model with the same 
assignments $(\Phi_e)^i_j$ and $(\Phi_u)^i_j$.
However, since we want to inherit the Sumino mechanism 
\cite{Sumino09} for the charged 
lepton mass relation, we adopt the assignment 
$\Phi_e^{i \alpha}$. 
On the other hand, we do not adopt 
the assignment $\Phi_u^{i \alpha}$, 
because if we adopt such the assignment, we are forced 
to modify the structure of $W'_u$ given in Eq.(3.13)
and $W_R$ given in Eq.(3.18) into more complicated 
forms in order to express these terms without $\Lambda$.

\begin{table}
\begin{center}
\begin{tabular}{|c|ccccccccccc|} \hline
   & $\bar{\bf 5}_i$ & ${\bf 10}_i$ & ${\bf 1}_i$ & 
$\bar{\bf 5}'_i$  & ${\bf 5}^{\prime\, i}$ & 
$\bar{\bf 5}^{\prime\prime\, i}$ & ${\bf 10}^{\prime\prime\, i}$ &
${\bf 5}^{\prime\prime}_i$ & $\overline{\bf 10}^{\prime\prime}_i$ &
$\bar{\bf 5}_H$ & ${\bf 5}_H$  \\ \hline
SU(5) & ${\bf 5}^*$ & ${\bf 10}$ & ${\bf 1}$ & 
${\bf 5}^*$ & ${\bf 5}$ & ${\bf 5}^*$ & ${\bf 10}$ & 
${\bf 5}$ & ${\bf 10}^*$ & ${\bf 5}^*$ & ${\bf 5}$  \\
SU(3) & ${\bf 3}$ & ${\bf 3}$ & ${\bf 3}$ & 
${\bf 3}$ &  ${\bf 3}^*$ & ${\bf 3}^*$ & ${\bf 3}^*$ & 
${\bf 3}$ & ${\bf 3}$ & ${\bf 1}$ & ${\bf 1}$ \\
O(3) & ${\bf 1}$ & ${\bf 1}$ & ${\bf 1}$ & ${\bf 1}$ &
 ${\bf 1}$ & ${\bf 1}$ & ${\bf 1}$ & ${\bf 1}$ & 
${\bf 1}$ & ${\bf 1}$ & ${\bf 1}$  \\
 \hline
\end{tabular}

\begin{tabular}{|cccccccccc|} \hline
 $\Sigma_3$ & $\Sigma_2$ & $Y_e$ & $Y_d$  & $Y_u$ & $Y_R$ & 
$\Phi_e$ & $\Phi_u$ & $\Omega_E$ & $\Omega_X$ 
\\ \hline
${\bf 24}+ {\bf 1}$ & ${\bf 24}+ {\bf 1}$ & ${\bf 1}$ & ${\bf 1}$ & 
${\bf 1}$ & ${\bf 1}$ & ${\bf 1}$ &  ${\bf 1}$ & ${\bf 1}$ & ${\bf 1}$ \\
${\bf 1}$ & ${\bf 1}$ & 
${\bf 6}^*$ & ${\bf 6}^*$ & ${\bf 6}^*$ & ${\bf 6}^*$ & 
${\bf 3}^*$ & ${\bf 8}+ {\bf 1}$ & ${\bf 3}^*$ & ${\bf 3}^*$  \\
 ${\bf 1}$  & ${\bf 1}$ &  
${\bf 1}$ & ${\bf 1}$ & ${\bf 1}$ & ${\bf 1}$ & ${\bf 3}$ & 
${\bf 1}$ & ${\bf 3}$ & ${\bf 3}$  \\
\hline
\end{tabular}

\begin{tabular}{|ccccccccccc|} \hline
${E}^u$ & ${P}_u$ &  $E$ & $X$ & 
$\Theta^e$ & $\Theta_u$ & $\Theta^{u \prime}$ & $\Theta^d$ &
$\Theta^E$ & $\Theta^X$ & $\Theta^R$
\\ \hline
 ${\bf 1}$ & ${\bf 1}$ & ${\bf 1}$ & ${\bf 1}$ & ${\bf 1}$ & ${\bf 1}$ &
${\bf 1}$ & ${\bf 1}$ & ${\bf 1}$ & ${\bf 1}$  & ${\bf 1}$ \\
${\bf 6}$ & ${\bf 6}^*$ & ${\bf 1}$ & ${\bf 1}$ &
${\bf 6}^*$ & ${\bf 8}+ {\bf 1}$ & ${\bf 6}$ & ${\bf 6}$ & 
${\bf 3}$ & ${\bf 3}$ & ${\bf 6}$ \\
${\bf 1}$ & ${\bf 1}$ &
${\bf 5}+ {\bf 1}$ & ${\bf 5}+ {\bf 1}$ & 
${\bf 1}$ & ${\bf 1}$ & ${\bf 1}$ & ${\bf 1}$ & 
${\bf 3}$ & ${\bf 3}$ & ${\bf 1}$ \\
\hline
\end{tabular}

\end{center}
\begin{quotation}
\caption{
Fields in the present model and their SU(5)$\times$U(3)$\times$O(3)
assignments. 
}
\end{quotation}
\end{table}


In Table 2, we list assignments of SU(5)$\times$U(3)$\times$O(3)
for all fields in the present model. 
The model is anomaly free in SU(5), while it is not so in the 
U(3) gauge symmetry.
Since the anomaly coefficients of ${\bf 3}$, ${\bf 6}$ and ${\bf 8}$ 
of SU(3) are $A({\bf 3})=1$, $A({\bf 6})=7$ and $A({\bf 8})=0$,
the sum of the anomaly coefficients $A$ is 
$\sum A=(42-29)A({\bf 3})+(5-5) A({\bf 6}) = 13$. 
In order that the model is anomaly free for U(3) family symmetry, 
we need further fields which give $\sum A=-13$, 
so that we may consider, 
for example, $A^{ij}$, $B^{i\alpha}$ and $C^{i\alpha}$.   
However, for the time being, we do not specify the roles of those 
fields $A$, $B$ and $C$ in the model.

$R$ charges of these fields in the yukawaon sector must satisfy
the following relations:
$$
R(Y_e) = 2 R(\Phi_e) \equiv r_e ,
\eqno(3.22)
$$
$$
R(Y_u) = 2 R(\Phi_u) -R(E^u) \equiv r_u ,
\eqno(3.23)
$$
$$
R(Y_d) = R(\Phi_u)+R(P_u) = 2R(\Omega_X)=2R(\Omega_E) 
= 2 \left( r_e +\frac{2}{3}\right) ,
\eqno(3.24)
$$
$$
R(Y_R) =R(\Phi_u) +R(Y_e) \equiv r_R .
\eqno(3.25)
$$
In these assignments, since we have still free parameters, 
we do not give numerical assignments of $R$ charges 
for these fields.
 
Finally, we would like to comment on $R$ parity assignments.
Since we inherit $R$ parity assignments in the standard SUSY model,
$R$ parities of yukawaons $Y_f$ (and also $\Theta_f$, $\Phi_{e,u}$, 
$E$, $\cdots$) are the same as those of Higgs particles 
(i.e. $P_R({\rm fermion})=-1$ and $P_R({\rm scalar})=+1$), 
while $(\bar{\bf 5}^{\prime\prime} + {\bf 5}^{\prime\prime})$,
$({\bf 10}^{\prime\prime} + \overline{\bf 10}^{\prime\prime})$, 
$\cdots$ are assigned to quark and lepton type, i.e. 
$P_R({\rm fermion})=+1$ and $P_R({\rm scalar})=-1$.

\vspace{5mm}

{\large\bf 4. Energy scales}

Masses of the additional particles $f'=(\bar{\bf 5}'+{\bf 5}')$ 
and $f^{\prime\prime}=(\bar{\bf 5}^{\prime\prime}+{\bf 10}^{\prime\prime})
+ ({\bf 5}^{\prime\prime}+\overline{\bf 10}^{\prime\prime})$ are 
given by 
$$
m(f') \sim \langle \Sigma_{2,3} \rangle, \ \ \ 
m(f^{\prime\prime}) \sim M_{5,10} .
\eqno(4.1)
$$
The existence of these additional particles does not
affect the value of $\Lambda_{GUT}$ in the minimal supersymmetric 
standard model (MSSM), so that the scale of $\Lambda_{GUT}$ is still 
given by
$$ 
\Lambda_{GUT} = 2\times 10^{16}\ {\rm GeV} .
\eqno(4.2)
$$
Since the VEV forms of $\Sigma_{2,3}$ break SU(5), it seems to be 
natural to consider 
$$
\langle \Sigma_{2} \rangle \sim \langle \Sigma_{3} \rangle 
\sim \Lambda_{GUT} 
\eqno(4.3)
$$ 
On the other hand, if we consider a lower value of $M_{5,10}$,
the gauge coupling constants $\alpha_3(\mu)$ will blow up 
before $\mu$ reaches the GUT scale $\Lambda_{GUT}$.
We have a constraint   
$$
M_{5,10}  \ge 10^{12} \ {\rm GeV}.
\eqno(4.4)
$$

Masses of the quarks and leptons are given as
$$
M_u =y_u\, y_{10}  \frac{\langle Y_u\rangle}{\bar{M}_{10}} \langle H_u^0 \rangle , 
\ \ \ \ 
M_{e,d} = y_{e,d}\, y_5 \frac{\langle Y_{e,d} \rangle}{\bar{M}_{5}} 
\langle H_d^0 \rangle , 
\eqno(4.5)
$$
from Eqs.(2.7) and (2.8), so that we have constraints 
$$
{\langle Y_u\rangle}/ {\bar{M}_{10}} \sim 1 , \ \ \ \ 
{\langle Y_e\rangle} / {\bar{M}_{5}} \sim 
{\langle Y_d\rangle} / {\bar{M}_{5}}  \sim 10^{-1}  , 
\eqno(4.6)
$$
which means
$$
(\bar{M}_{10})^2 \sim (M_{10})^2 \sim {\langle Y_u\rangle}^2 , \ \ \ \ 
(\bar{M}_{5})^2 \sim (M_{5})^2 \gg {\langle Y_{e,d}\rangle}^2  , 
\eqno(4.7)
$$
respectively.

Taking the constraints (2.17) and (4.4) into consideration, 
we assume that particles $f^{\prime\prime}$ have 
masses of the order of 
$$
M_5 \sim M_{10} \sim 10^{14} \ {\rm GeV} ,
\eqno(4.8)
$$ 
so that an energy scale at which U(3) is broken is  
$$
\langle Y_f \rangle \sim  
\langle Y_R \rangle \sim \Lambda_{fam}
\sim 10^{13} \ {\rm GeV}.
\eqno(4.9)
$$
For reference, we illustrate the behaviors of the gauge coupling 
constants $\alpha_i(\mu)$ ($i=1,2,3$) for the 
case $M_{5,10} = 10^{14}$ GeV in Fig.1.

\begin{figure}[t!]
  \includegraphics[width=60mm,clip]{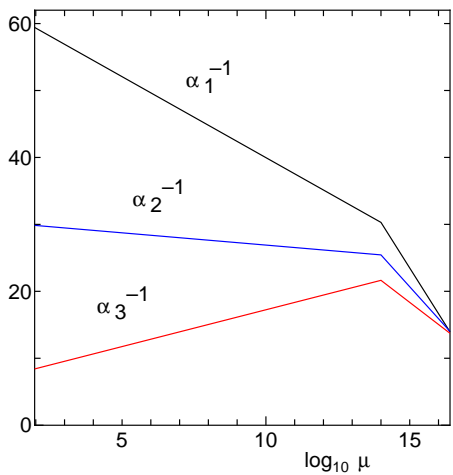}
  \caption{ Behaviors of gauge coupling constants $\alpha_i^{-1}$
($i=1,2,3$) in the case of $M_{5,10}=10^{14}$ GeV.
For simplicity, we have neglected the SUSY breaking effects at 
$\mu \sim 10^3$ GeV in this figure. 
}
  \label{all}
\end{figure}

As seen in Table 2, we have many U(3) non-singlet fields 
in the present model, so that the model does not give an 
asymptotic free theory.
The gauge coupling constants of U(3) will blow up before 
$\mu$ reaches $\mu=\Lambda_{GUT}$ 
if we take a lower value of $\Lambda_{fam}$.
In this paper, we adopt a scenario based on Eq.(4.9). 
Then, since the scale $M_{5,10}$ is very high, the family gauge
coupling constants $\alpha_{fam}$ does not blow up from 
$\mu=M_{5,10}$ up to $\mu=\Lambda_{GUT}$.
 
We do not discuss the behaviors of gauge coupling constants 
above $\mu=\Lambda_{GUT}$ because we have no scenario at 
$\mu > \Lambda_{GUT}$ at present.

\vspace{5mm}

{\large\bf 5. Concluding Remarks}

In conclusion, we have investigated a yukawaon model which is 
compatible with an SU(5) GUT scenario. 
Since yukawaons are SU(5) singlets, the existence of the yukawaons
do not affect the SU(5) GUT model, so that we can inherit the successful 
results in the SU(5) GUT and we also inherit the current problems in the 
minimum SU(5) GUT scenario.
We optimistically consider that those problems peculiar to the SU(5) scenario
will be resolved in a model based on a higher GUT symmetry and/or 
on extra-dimension scenario.

In the present model, we have the following matter fields:
$$
(\bar{\bf 5}+{\bf 10}+ {\bf 1})_i + (\bar{\bf 5}'_i+{\bf 5}^{\prime\, i})
+(\bar{\bf 5}^{\prime\prime}+{\bf 10}^{\prime\prime})^i 
+({\bf 5}^{\prime\prime}+\overline{\bf 10}^{\prime\prime})_i .
\eqno(5.1)
$$
We do not consider ${\bf 1}_i^{\prime\prime}+{\bf 1}^{\prime\prime\, i}$
The particles $f'=(\bar{\bf 5}'_i+{\bf 5}^{\prime\, i})$ and
$f^{\prime\prime} = (\bar{\bf 5}^{\prime\prime}+{\bf 10}^{\prime\prime})^i 
+({\bf 5}^{\prime\prime}+\overline{\bf 10}^{\prime\prime})_i$ 
have masses of the orders of $\Lambda_{GUT} \sim 10^{16}$ GeV and
$M_{5,10} \sim 10^{14}$ GeV, respectively.
The U(3) family symmetry is broken at $\mu= \Lambda_{fam} 
\sim 10^{13}$ GeV, 
whose value has been settled by a neutrino seesaw mass
and no blowing up condition of the conventional gauge 
coupling constants. 

The most notable result in the present SU(5) compatible model is
that the model naturally lead to a model without $Y_\nu$ in which 
$Y_e$ plays a role of a substitute for $Y_\nu$, 
although it was an ac hoc assumption in the previous
yukawaon model \cite{O3_PLB09}. 

It is also worthwhile to notice that the present model is
a model without a cutoff scale $\Lambda$ differently from
the past yukawaon models, and the factors $\bar{M}_{10}$ and
$\bar{M}_{5}$ in the seesaw expressions (2.7) and (2.8) 
have family-dependence. 
In the past yukawaon models, since the model have been based 
on an effective theory with a cutoff parameter $\Lambda$, 
we are obliged to consider $\langle Y_u \rangle/\Lambda \sim 1$
in order to explain the observed fact $m_t \sim \langle H_u^0\rangle$.
However, in the effective theory with the scale $\Lambda$, it was
unnatural to consider $\langle Y_u \rangle \sim \Lambda$.
In the present model, since the cutoff parameter $\Lambda$ is
replaced with $\bar{M}_{10}=M_{10}(1+\langle Y_u \rangle^2/M_{10}^2)$,
we can safely choose $\langle Y_u \rangle \sim M_{10}$.
Besides, we can expect visible effects in the phenomenology of 
the up-quark mass matrix due to $\bar{M}_{10} \neq M_{10}$, 
although the effects due to $\bar{M}_5 \neq M_5$ in the
charged lepton and down-quark sectors are negligibly small.
Phenomenological investigation based on the present model
with $\bar{M}_{10} \neq M_{10}$ will be given elsewhere.

Regrettably, we have failed to build a model
in which U(3) family gauge boson effects are visible (for instance, 
Ref.\cite{KSY_PLB11}).
It seems to be hard to embed a lower scale of $\Lambda_{fam}$
(e.g. $\Lambda_{fam} \sim 10^{7-8}$ GeV \cite{U3O3_JPG11})
 into the present SU(5) compatible yukawaon model.
This is still our future task.

\vspace{5mm}
{\Large\bf Acknowledgments}   

The author would like to thank T.~Yamashita for helpful conversations. 
The work is supported by JSPS 
(No.\ 21540266).

%



\begin{thebibliography}{99} 
%

\bibitem{yukawaon} Y.~Koide, Phys.~Rev. {\bf D 79}, 033009 (2009).
%
\bibitem{flavon}  C.~D.~Froggatt and H.~B.~Nelsen, Nucl.~Phys. 
{\bf B 147}, 277 (1979).
%
\bibitem{O3_PLB09}
Y.~Koide, Phys.~Lett. {\bf B 680}, 76 (2009).
%
%
\bibitem{SU5} H.~Georgi and S.~L.~Glashow, Phys.Rev.Lett. 
{\bf 32}, 438 (1974). 
%
\bibitem{T-D_splitting} S. Dimopoulos and F. Wilczek, 
in {\it The Unity of the Fundamental Interactions}, Proceedings
of the 19th Course of the International School of Subnuclear Physics, 
Erice, Italy, 1981, edited by A. Zichichi (Plenum Press, New York, 1983);
M. Srednicki, Nucl. Phys. B202, 327 (1982).
%
\bibitem{PDG10} Particle Data Group, K.~Nakamura, {\it et al}., 
J.~Phys. {\bf G 37}, 075021 (2010).
%
%
\bibitem{tribi} P. F. Harrison, D.~H.~Perkins and W.~G.~Scott, 
Phys.~Lett. {\bf B 458}, 79 (1999); 
Phys.~Lett. {\bf B 530}, 167 (2002); 
Z.-z.~Xing, Phys.~Lett. {\bf B 533}, 85 (2002); 
P.~F.~Harrison and W.~G.~Scott, Phys.~Lett. {\bf B 535}, 163 (2002);
 Phys.~Lett. {\bf B 557}, 76 (2003); 
E.~Ma, Phys. Rev. Lett. {\bf 90}, 221802 (2003); 
C.~I.~Low and R.~R.~Volkas, Phys.~Rev. {\bf D 68}, 033007 (2003).
%
%
\bibitem{Sumino09} Y.~Sumino,  Phys.~Lett. {\bf B671}, 477 (2009);
JHEP {\bf 0905}, 075 (2009). 
%
\bibitem{Koide90MPL}  Y.~Koide, Mod.~Phys.~Lett. {\bf A5}, 2319 (1990). 
%
\bibitem{Koidemass} Y.~Koide, Lett.~Nuovo Cimento {\bf 34}, 201 
(1982); Phys.~Lett. {\bf B120}, 161 (1983);
Phys.~Rev. {\bf D28}, 252 (1983).
%
\bibitem{Arason} H.~Arason, {\it et al.}, Phys.~Rev. 
{\bf D 46}, 3945 (1992).
%
\bibitem{Appelquist06} T.~Applequist, Y.~Bai and M.~Piai, Phys.~Rev.
{\bf D 74},  076001 (2006); Phys.~Lett. {\bf B 637}, 245 (2006).
%
\bibitem{U3O3_JPG11} Y.~Koide, J.~Phys. {\bf G 38}, 085004 (2011). 
%
%
%
\bibitem{q-mass} Z.-z.~Xing, H.~Zhang and S.~Zhou, 
{Phys.~Rev.} {\bf D 77}, 113016 (2008).
And also see, H.~Fusaoka and Y.~Koide, {Phys. Rev.} 
{\bf D 57}, 3986 (1998).
%
\bibitem{KSY_PLB11} Y.~Koide, Y.~Sumino and M.~Yamanaka, 
Phys.~Lett. {\bf B695}, 279 (2011).
%
\end{thebibliography}
\end{document}